\documentclass{elsart}

\usepackage{graphicx}

\usepackage{amssymb}

\begin{document}

\begin{frontmatter}



\title{Memory distribution in complex fitness landscapes}


\author{Juan G. Diaz Ochoa}

\address{Fachbereich 1, University of Bremen, Otto Hahn Allee, D-28359 Bremen, Germany}

\begin{abstract}
In a co-evolutionary context, the survive probability of individual elements of a system depends on their relation with their neighbors. The natural selection process depends on the whole population, which is determined by local events between individuals. Particular characteristics assigned to each individual, as larger memory, usually improve the individual fitness, but an agent possess also endogenous characteristics that induce to re-evaluate her fitness landscape and choose the best-suited kind of interaction, inducing a non absolute value of the outcomes of the interaction. In this work, a novel model with agents combining memory and rational choice is introduced, where individual choices in a complex fitness landscape induce changes in the distribution of the number of agents as a function of the time. In particular, the tail of this distribution is fat compared with distributions for agents interacting only with memory.
\end{abstract}


\end{frontmatter}

\section{Introduction}
The emergence of cooperation without an external enforcing agency, as an external coordinator, makes game theory based models attractive\cite{Smith}\cite{Hauert}\cite{Nowak}. Usually, this assumption is done on the basis that the interaction between elements can be absolutely defined, i.e., the fitness scale has an absolute value \cite{Nowak}. The prisoner's dilemma appears as a paradigm to model populations of interacting agents that are able to form clusters of cooperative individuals. What happen if the interaction between the elements is no more absolute? It is quite reasonable to assume non absolute values of fitness scales in interactions between agents. Different mechanisms can be responsible for this particular characteristic. Either the agents can suffer or induce changes in their interactions (e.g., in ecology by moving from one to other ecological niche) or they can simply avoid the interactions with other competitors. In both cases the induced changes are not trivially fortuitous. For instance, some species of non-migratory birds, that have an ecosystem at the river of a lake, are subjected to strong competition against other birds of the same species due that some places are overcrowded. During the reproductive phase this represent a high mortality rate. But it had been observed that this species of birds make habitat selection, moving from one habitat with high mortality rate to other place with higher fitness \cite{Morris}. This change of fitness compensates the mortality of the species.

A frequent problem in game theory is the constant measurement unit assigned to the outcomes of the game \cite{Hofbauer}. Even in bimatrix games, where two different elements have the option to interact with different interaction matrices, the measurement is absolute \cite{Sigmund_Nowak}\cite{Hofbauer}. If we suppose the system is perfect isolated, and the elements have available only a limited interaction palette, then the elements have no other choice than to dispose of a single simple interaction matrix. In contrast, if the elements are embedded in an open system, and additionally have an internal reservoir that allows the search in the configuration space, the interaction matrix cannot be absolutely defined. Therefore, we introduce a very simple improvement in this kind of models by means of the definition of flexible matrices together with individual bias (preferences) for each agent. In this context, the novel approach presented here is the implementation of relative values of the interaction matrix.

\section{Model}

In this model, the agents can remember actions in the past and they can take decisions according to their stored information. Simple learning schemas can be defined. For instance, one schema is 'copy the best', where elements try to copy the best interaction from their neighborhood and implement it in their next interaction \cite{Nowak_May}. Memory is modeled as strips of information stored in the memory of agent $i$ that are projected onto vectors of strategies in order to produce new actions represented by vectors ${\bf \sigma}^{i}$ \cite{Lindgren_I}.

The pair interaction is non commutative and is defined using an interaction matrix ${\bf F}$. The implemented game is a prisoner's dilemma defined by a $2 \times 2$ matrix, where the chain inequality $T > R > P > S$ is obeyed ('T' for temptation, 'R' for reward, 'P' for punishment and 'S' for sucker). In this model only pair interactions are allowed. The interaction for an agent and her adversary can be either cooperate -equivalent in statistical mechanics to spin up- or defect -equivalent to spin down-. The vector notes each strategy ${\bf\sigma}$. If  ${\bf \sigma}^{i}$ is the interaction of the agent $i$ and ${\bf \sigma}^{j}$ is the interaction of agent $j$, then, the total utility matrix for the agent $i$ is given by $U^{j i} = {\bf \sigma}^{j}{\bf F}{\bf \sigma}^{i}$, and the total utility of the agent $i$ is defined as $f^{i} = \sum_{j}U^{ij}$ \cite{Hofbauer}. Under the imposed constraints imposed by ${\bf F}$, it is rational for the two agents to defect. But also it would be of mutual advantage for the player to establish cooperation in a long run. In the model the interactions are implemented in such a way that the agents have perfect information from the past, but no information of the instantaneous interaction of the adversary.

The definition of non absolute values in the interactions is made by means of an extended interaction matrix ${\bf F_{p}}$. Furthermore, an additional vector ${\bf \zeta _{k}}$ defines the preferred interaction of the agent $i$. This vector points into each one of the interaction sub-matrixes to be fixed in each interaction. The utility matrix for the agent $i$ has in this case the following form
\begin{equation}
U^{ji}=\delta_{il}{\bf \sigma^{j}}[{\bf \zeta^{l}}{\bf F_{p}}{\bf \zeta^{i}}]{\bf \sigma^{i}}.
\end{equation}
The games are characterized by three kinds of sub-matrices. One sub-matrix is defined under the restriction $2R > T+S$. The second sub-matrix is defined as a chicken game, which means, the mutual defection is the worst possible outcome. Therefore, the values for this sub-matrix are $T'>R'>S'>P'$. The values were arbitrary defined, but in such a away that $T'< R'$, i.e., 'temptation' is not dramatically different to 'reward'. Additionally, $T'<T$, i.e. is a low risk game. A third sub-matrix representing a constant interaction is introduced, with the condition $T''=R''=S''=P''=1$. This last sub-matrix implies for the agents a total risk aversion, but low incomes. Hence, the agents interacting in the present model could basically choose between more or less risk. Simultaneously, these three fundamental options allow the game to switch between stable and unstable equilibrium states.

This model assumes a number of $N$ agents situated in a lattice (see fig. \ref{fig1}). Given that this is a square lattice, each element can only interact with its four nearest neighbors, according to the Von Neumann Schema. The memory size $M$ of each agent is the main observable of the system. The acceptance or rejection depends on the utility $f^{i}$ obtained by agent $i$. Naturally, the evolution of the memory size of the individuals automatically implies changes in the local strategies of the elements. At $t=0$ there is a random distribution of agents with low memories ($M^{i}<2$). If the total utility $f^{i}$ of the element $i$ is smaller than the utility of the adjunct opponents $j$, then a mutation of the memory size of $i$, $M^{i}$, occurs, i.e. $M^{i} \rightarrow (M+1)^{i}$. This mutation takes place at frequencies $\omega_{m}$, that are proportional to the memory size. The reason is, the assumption that the mutation probability in complex elements does not take place with the same frequency as the mutation in much simpler elements.

\section{Results}

\begin{figure}[ht]
\begin{center}
\includegraphics[clip, width=0.43\textwidth]{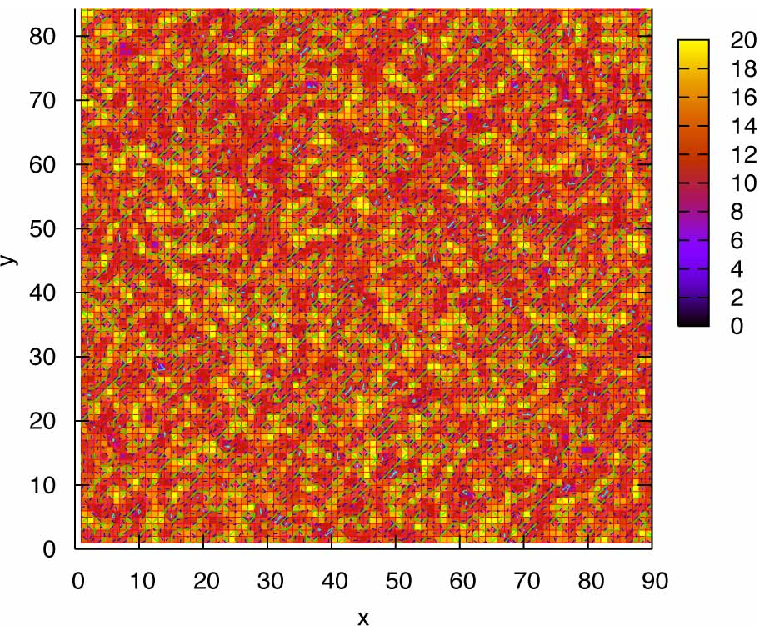}
\includegraphics[clip, width=0.40\textwidth]{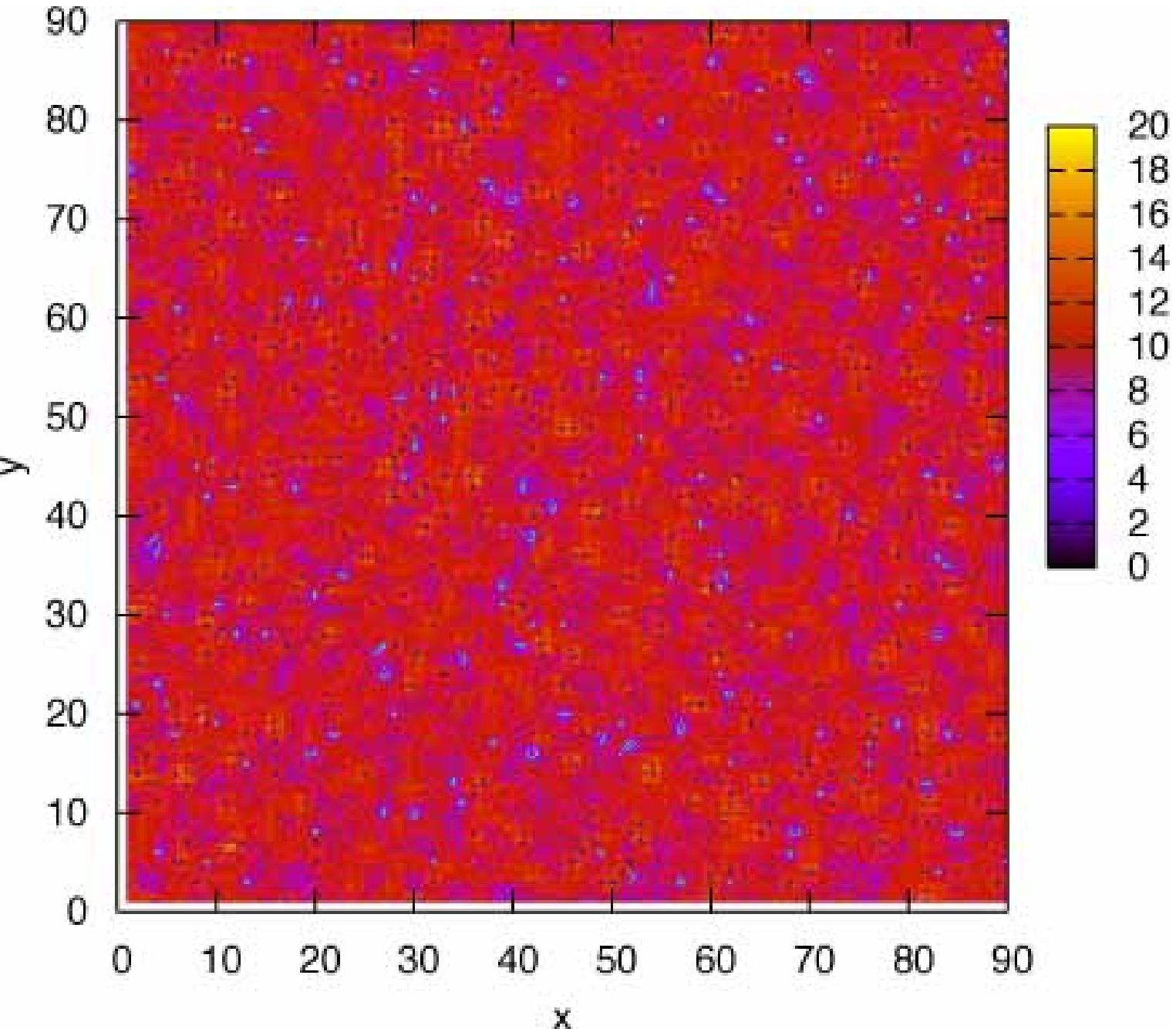}
\caption{Snapshot of the distribution of memories. The figure on the left side corresponds to prisoner's dilemma interactions. The right side figure represents mixed interactions and agents with individual choices.}
\label{fig1}
\end{center}
\end{figure}

We make a qualitative analysis of the distribution of memories when the agents have no preferences, i.e., when the interactions are given by the canonical matrix $F=(3,0,5,2)$. After a long simulation time a relative homogeneous mixture of elements with low memory suffers a transition. The mixture becomes inhomogeneous and forms a structure that alternates between different memory components. The colors in fig. \ref{fig1} symbolize different memory size. Blue corresponds to agents with low memories, whereas yellow symbolizes agents with high memory. After long iteration time non-symmetric patterns emerge. Although these patterns do not have a clear structure, they are also not completely random. We compare this result with the behavior of agents allowed to choose into the flexible fitness landscape (right side figure). At the same time scale the distribution of agents with low memory dominates over the agents with high memory, and only few non symmetric patterns with high memory agents appears.
The time dependence of the distribution of memory shows a rapid growth and subsequent extinction after some characteristic time. The mechanism is in such case very simple: if the fitness of the agents with memory $M$ is large enough, then they do not find any resistance and can replace other elements in the neighborhood and spread into the system. If a new class of elements with a much better fitness and other memory size $M'$ appears, these new elements start to spread. Eventually the agents with less fitness, i.e. agents with memory $M$, start an extinction process. The extinction rate strongly depends on the interaction each agent chooses. When the agents interact with flexible matrices $<F>$, which is equivalent to agents with choices into the extended fitness landscape ${\bf F_{p}}$, the fluctuation-dissipation is no more in equilibrium. With less dissipation the agents increase their chance to reproduce themselves. This fact is translated in the existence of self-similar processes that take place in different time regimes. Therefore, the distribution in time of the number of agents with memory $M$, $N[M]$, have a fat tail when the interactions are more flexible than the simple interaction governed by the prisoner's dilemma.
\begin{figure}[ht]
\begin{center}
\includegraphics[clip, width=0.60\textwidth]{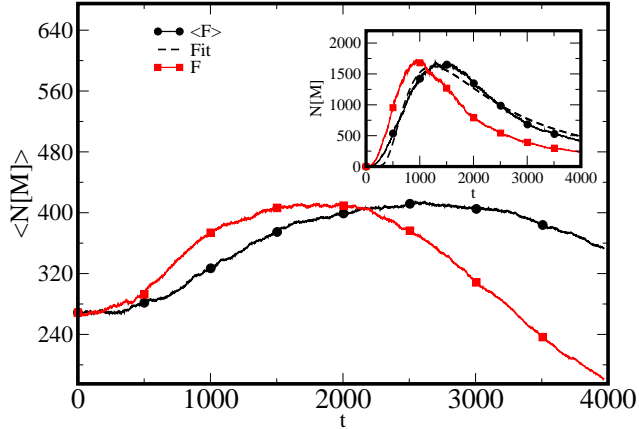}
\caption{Distribution in time of the total population of the system $<N[M]>$ as a function of the time for flexible interactions $<F>$ (agents with choices into the extended fitness landscape ${\bf F_{p}}$) and prisoner's dilemma ${\bf F}$.. The inset shows the non-normalized distribution of the number of agents with memory $M=6$ and fit.}
\label{fig2}
\end{center}
\end{figure}
We can then concentrate us in a specific memory. The distribution function is asymmetric and can be fitted with the following distribution \cite{paul}
\begin{equation}
N[M](\tau) = \sqrt{ \frac{A_{0}[M]}{2\pi}}\frac{e^{-A_{0}[M]/2\tau}}{\tau^{5/2}},  
\end{equation} 
where $A_{0}[M]$ is a fit parameter for each memory size. Given that the fit is valid for a long time scale, the time parameter $\tau$ is defined as $\tau = \frac{t}{T_{0}}$, where $T_{0}$ is the characteristic frequency where a growth and subsequent extinction of the referred memory size takes place. In the present computations $T_{0} \sim T$, where $T$ is the period where the computations were done. This fit is a reasonable approximation to the shape of the density function and is shown in the inset of Fig. \ref{fig2}. In the same figure, the average of the population of different memories $<N[M]>$ is shown. In long time regimes the suppression of the dissipation in the case of flexible interactions allows a relative stability in the total population (more diversity), whereas for the simple prisoner's dilemma a mass extinction takes place.
\section{Sumary}
The present work describes quantitatively the ability that an agent has to choose in a complex energy landscape. This characteristic of the agents cannot simply be ignored and must be implemented including individual memory features.

To sum up, the present results are a heuristic approximation to describe systems consisting of agents that can modify their fitness. Two main aspects can be extracted from the present model: first, the possibility to obtain communities with high diversity and inhomogeneous spatial distribution. Second, this diversity depends on the ability that each agent shows to choose the kind of interaction she needs.

I want to thank H. Fort for very useful comments. 


\begin{thebibliography}{}

\bibitem{Smith} M. Smith, \emph{Evolution and the theory of games}, Cambridge University Press (1998).

\bibitem{Hauert} C. Hauert, G Szab\'{o}, Am. J. Phys. {\bf 73}, 406, (2005).  

\bibitem{Nowak} M. A. Nowak and K. Sigmund \emph{Science}, {\bf 303}, 793 (2004).

\bibitem{Morris} D. W. Morris, \emph{Nature}, {\bf 443}, 645 (2006).

\bibitem{Hofbauer} J. Hofbauer and K. Sigmund, \emph{Evolutionary games and Population Dynamics}, Cambridge University Press (1998).

\bibitem{Sigmund_Nowak} K. Sigmund, M. Nowak, Curr. Biol., R504.

\bibitem{Nowak_May} M.A. Nowak and R.M. May \emph{Nature}, {\bf 359}, 826, (1992).

\bibitem{Lindgren_I} K. Lindgren, in: \emph{Artificial life II}, C.G. Lagnton, J.D. Farmer, S. Rasmunsen and C. Taylor eds., Addyson Wesley, (1991).

\bibitem{paul} W. Paul, J. Baschnagel, \emph{Stochastic processes: from physics to finance}, Springer (1999).

\end{thebibliography}
\end{document}